\def\thisversion{27 June 2024}
\documentclass[aps,prd,twocolumn,%
                nofootinbib,showpacs]{revtex4-2}

\usepackage{amsmath,amssymb}
\usepackage{mathrsfs}

\newcommand*{\EMT}{\mathcal{T}}
\newcommand*{\mscrL}{\mathscr{L}}
\newcommand*{\op}[1]{\textsf{#1}}

\allowdisplaybreaks

\begin{document}

\title{Using gauge invariance to symmetrize\\ the energy-momentum tensor of electrodynamics}

\author{Helmut Haberzettl}
 \email{helmut@gwu.edu}
 \affiliation{GW Institute for Nuclear Studies and Department of Physics, The George Washington University, Washington, DC 20052,
 USA}

\date{\thisversion}

\begin{abstract}
It is shown that using Noether's Theorem explicitly employing gauge
invariance for variations of the electromagnetic four-potential $A^\mu$
straightforwardly ensures that the resulting electromagnetic energy-momentum
tensor is symmetric. The Belinfante symmetrization procedure is not
necessary. The method is based on Bessel-Hagen's 1921 clarification of
Noether's original procedure, suggesting that the symmetry problem arises
from an incomplete implementation of Noether's Theorem. The derivation
addresses in some detail where the usual application of Noether's Theorem
falls short, what the Belinfante procedure actually does to fix the problem,
and why the usual unsymmetric canonical energy-momentum tensor can only be
used for extracting four-momentum conservation based on translational
invariance, but will provide meaningless results when applied to  rotations
or boosts, unless modified appropriately.
\end{abstract}

\maketitle


\section{Introduction}  \label{sec:introduction}

It is a well-known problem that the standard application of Noether's
Theorem~\cite{noether} to the Lagrangian density for the free electromagnetic
field\footnote{%
   We use SI units and the Minkowski metric $\text{diag}(+1,-1,-1,-1)$.
  }
\begin{equation}
  \mscrL= -\frac{1}{4\mu_0} F^{\mu\nu}F_{\mu\nu}
  \label{eq:Lagrangian}
\end{equation}
produces an energy-momentum tensor that is not
symmetric~\cite{kaku,peskin,jackson}. The field-strength tensor appearing here,
\begin{equation}
  F^{\mu\nu} =\partial^\mu A^\nu-\partial^\nu A^\mu~,
  \label{eq:fieldstrength}
\end{equation}
is the antisymmetric four-curl of the electromagnetic four-potential $A^\mu$.
To account for the ten independent parameters of the Poincar\'e group, it is
necessary and sufficient, and thus essential, that the energy-momentum tensor
be symmetric (with ten independent elements). To achieve this, the unsymmetric
tensor is subjected to the symmetrization procedure proposed by
Belinfante~\cite{belinfante} (see also \cite{rosenfeld}), resulting in a
symmetric energy-momentum tensor, capable of producing all conserved entities
of electrodynamics.

Given the fundamental nature of Noether's Theorem, it seems quite inconceivable
that it would not be capable of producing a symmetric energy-momentum tensor as
a matter of course, without the necessity for additional procedures. Indeed, it
was clarified by Bessel-Hagen~\cite{BesselHagen} over a century ago that making
use of the full set of applicable symmetries will produce a symmetric
energy-momentum tensor in a straightforward manner. Unfortunately,
Bessel-Hagen's work is not very well known and textbooks still present the
Belinfante symmetrization procedure as the standard way of treating the
problem. Bessel-Hagen's work clarifies that the variational procedure
underlying Noether's Theorem is not restricted to the usual spacetime and
functional variations, the way it is usually interpreted, but also must include
mixtures of spacetime and functional variations, like the gauge-invariance
considerations to be discussed below, to extract the full dynamical symmetries
of a problem.

We will discuss here the application of the Bessel-Hagen procedure to the free
electromagnetic Lagrangian (\ref{eq:Lagrangian}). Some of the results to be
provided here --- even though not widely known --- are not unknown
\cite{eriksen, munoz,burgess,montesinos,scheck,BLS,hobson}, but to our
knowledge have never been presented in a manner that specifically addresses the
shortcomings of the usual variational approach to electrodynamics as found in
textbooks. The present note is intended to fill this gap.

To discuss the problems with the standard approach, it will be necessary to
recap some of the details of the variational formalism underlying Noether's
Theorem. This will show, in particular, that the usual textbook derivation of
the so-called canonical (i.e., unsymmetric) energy-momentum tensor is flawed
because it ignores infinitesimal variations related to rotation and boost
degrees of freedom (which then, not surprisingly, lead to the corresponding
well-known problems with the canonical energy-momentum
tensor~\cite{kaku,peskin,jackson}). Explicitly incorporating gauge invariance
in the application of Noether's Theorem will show that a mixture of
infinitesimal spacetime and gauge-freedom variations will compensate for this
shortcoming and provide a gauge-invariant, symmetric energy-momentum tensor as
a matter of course capable of describing all conservation laws of
electrodynamics.


\section{Variational Approach}

To set the stage, let us recap some details of the variational procedure
underlying Noether's Theorem. Applied to the electromagnetic fields, one
considers the invariance of the action integral
\begin{equation}
  S= \int_R d^4x\, \mscrL(A^\nu,\partial^\mu A^\nu,x^\mu)
\end{equation}
under variations with fixed end points,
\begin{subequations}\label{eq:xAvariations}
\begin{align}
  x'^\mu &= x^\mu+\delta x^\mu~,
  \\
  A'^\mu(x) &= A^\mu(x) + \delta \bar{A}^\mu(x)~,
  \label{eq:barA}
\end{align}
\end{subequations}
where $\delta x^\mu$ is the variation of the spacetime variable $x^\mu$ and
$\delta \bar{A}^\mu(x)$ is the variation of the four-potential $A^\mu$. More
details of $\delta \bar{A}^\mu(x)$ will be discussed below. The Lagrangian
density $\mscrL$ here must be a Lorentz scalar and $R$ is a simply connected
four-dimensional spacetime region bounded by a spacelike hypersurface $\partial
R$ on which variations vanish. The usual variational procedures (assumed to be
known; see \cite{gelfand}) then show that the vanishing variation of the
action, $\delta S=0$, is equivalent to
\begin{align}
  \delta S
  &=\int_{R} d^4x\left[\frac{\partial \mscrL}{\partial A^\nu}
  -\partial^\lambda \frac{\partial \mscrL}{\partial(\partial^\lambda A^\nu ) } \right]\delta\bar{A}^\nu
  \nonumber\\
  &\qquad\mbox{}+\int_{R} d^4x \,\partial^\mu
  \left[\frac{\partial \mscrL}{\partial(\partial^\mu A^\lambda ) } \delta\bar{A}^\lambda
  +\mscrL \delta x_\mu \right]
  \nonumber\\
  &=\int_{R} d^4x\left[\frac{\partial \mscrL}{\partial A^\nu}
  -\partial^\lambda \frac{\partial \mscrL}{\partial(\partial^\lambda A^\nu ) } \right]\delta\bar{A}^\nu
  \nonumber\\
  &\qquad\mbox{}+\oint_{\partial R} d\sigma^\mu
  \left[\frac{\partial \mscrL}{\partial(\partial^\mu A^\lambda ) } \delta\bar{A}^\lambda
  +\mscrL \delta x_\mu \right]=0~,
  \label{eq:Noether1}
\end{align}
where the volume integral with an overall divergence was converted into a
hypersurface integral over $\partial R$ (with spacelike three-dimensional
surface element $d\sigma^\mu$) with the help of Gauss's theorem. The surface
integral vanishes since, by construction, variations vanish on $\partial R$.
The independence of the variations $\delta\bar{A}^\nu$ then produce the
Euler--Lagrange equations,
\begin{equation}
\frac{\partial \mscrL}{\partial A^\nu}
  -\partial^\lambda \frac{\partial \mscrL}{\partial(\partial^\lambda A^\nu ) }=0~,
  \label{eq:ELeq}
\end{equation}
and the surface integral vanishes separately,
\begin{equation}
\oint_{\partial R} d\sigma^\mu
  \left[\frac{\partial \mscrL}{\partial(\partial^\mu A^\lambda ) } \delta\bar{A}^\lambda
  +\mscrL \delta x_\mu \right]=0~.
  \label{eq:Noether0}
\end{equation}
Exploiting this result is the basis for Noether's Theorem~\cite{noether}. To
evaluate it further, we may write
\begin{equation}
  \delta\bar{A}^\lambda = \big[A'^\lambda(x)-A'^\lambda(x')\big]+ \big[A'^\lambda(x')-A^\lambda(x)\big]~.
  \label{eq:dAsplit}
\end{equation}
The second bracket produces
\begin{equation}
A'^\lambda(x')-A^\lambda(x)
=\frac{\partial x'^\lambda}{\partial x^\sigma} A^\sigma(x) - A^\lambda(x)
\label{eq:Anutransformed}
\end{equation}
since $A^\lambda$ transforms like a contravariant vector field, with
\begin{equation}
\frac{\partial x'^\lambda}{\partial x^\sigma}
    =\delta^\lambda_{\sigma} + \partial_\sigma (\delta x^\lambda)~,
\end{equation}
where $\delta x^\lambda$ is an infinitesimal Lorentz transformation of the
spacetime variable $x^\lambda$. If the transformation is a simple translation,
the derivative $\partial_\sigma (\delta x^\lambda)$ vanishes, but if it is a
rotation or boost, it does not. One then obtains
\begin{equation}
\partial_\sigma (\delta x^\lambda) = \partial_\sigma \op{X}^{\lambda\rho\tau} \omega_{\rho\tau}~,
\end{equation}
where $\omega_{\rho\tau}$ is the usual antisymmetric matrix containing the
infinitesimal boost and rotation parameters~\cite{kaku,peskin,jackson}. The
details are unimportant other than the fact that its derivative vanishes.
However, the tensor
\begin{equation}
\op{X}^{\lambda\rho\tau} = g^{\lambda\rho}x^\tau - g^{\lambda\tau} x^\rho~,
\end{equation}
which is the generator of rotations and boosts, has a non-vanishing derivative.
We will refer to such contributions simply as rotational to avoid long-winded
expressions. Hence, we obtain
\begin{equation}
 A'^\lambda(x')-A^\lambda(x)
 = A^\sigma (\partial_\sigma \op{X}^{\lambda\rho\tau}) \omega_{\rho\tau}\equiv \delta_1 A^\lambda~,
 \label{eq:delta1}
\end{equation}
which accounts for the non-vanishing contribution under infinitesimal rotations
if $\omega_{\rho\tau}\neq0$. The notation $\delta_1$ here signifies that we
will encounter similar variational contributions later. The variation $\delta
\bar{A}^\lambda$ of Eq.~(\ref{eq:dAsplit}) now reads
\begin{equation}
\delta\bar{A}^\lambda = \left[A'^\lambda(x)-A'^\lambda(x')\right]  +  \delta_1 A^\lambda~.
\label{eq:dAprime}
\end{equation}
Accounting only for spacetime variations, to first order the bracket term
produces
\begin{equation}
A'^\lambda(x)-A'^\lambda(x')  = -(\partial_\nu A^\lambda) \delta x^\nu~.
\label{eq:deltaAcanonical}
\end{equation}
If we now, for the time being, ignore the infinitesimal rotation contribution
$\delta_1 A^\lambda$ in Eq.~(\ref{eq:dAprime}) and only use the term
(\ref{eq:deltaAcanonical}), the surface integral (\ref{eq:Noether0}) may then
be recast in the form
\begin{equation}
\oint_{\partial R} d\sigma_\mu
  \left[g^{\mu\sigma}\frac{\partial \mscrL}{\partial(\partial^\sigma A^\lambda ) } \partial^\nu A^\lambda
  -g^{\mu\nu}\mscrL\right] \delta x_\nu =0~
  \label{eq:Noether2}
\end{equation}
This is the standard textbook result with the usual unsymmetric energy-momentum
tensor --- often called the canonical energy-momentum tensor --- appearing here
in the square brackets,
\begin{align}
  T_c^{\mu\nu} &= g^{\mu\sigma}\frac{\partial \mscrL}{\partial(\partial^\sigma A^\lambda ) } \partial^\nu A^\lambda
  -g^{\mu\nu}\mscrL
  \nonumber\\
  &=-\frac{1}{\mu_0} F^\mu_{~\lambda} \partial^\nu A^\lambda-g^{\mu\nu}\mscrL~,
  \label{eq:EMTunsym}
\end{align}
where the index $c$ stands for canonical. It should be clear that this result
is only valid for translational degrees of freedom since possible rotational
contributions were dropped in its derivation.


\subsection{Effect of neglected rotational contributions}\label{sec:roteffect}

Omitting the rotational $\delta_1 A^\lambda$ contribution from the integral
(\ref{eq:Noether2}) is precisely the reason for the well-known
finding~\cite{kaku,peskin,jackson} that one cannot extract angular-momentum
properties from the tensor (\ref{eq:EMTunsym}) in any meaningful way since when
attempting to do so, one must write the spacetime increment in
(\ref{eq:Noether2}) in the form appropriate for infinitesimal rotations, namely
$\delta x_\nu = \op{X}_\nu^{~\rho\tau} \omega_{\rho\tau}$, which is of the same
form as $\delta_1 A^\lambda$.  Adding the omitted term back replaces the
expression (\ref{eq:Noether2}) by
\begin{align}
\oint_{\partial R} d\sigma_\mu\big(S^{\mu\rho\tau}
 \omega_{\rho\tau}
  -T_c^{\mu\nu} \delta x_\nu\big) &=0~,
  \label{eq:Noether5}
\end{align}
where the extra term depends on the so-called spin-angular-momentum tensor
given by
\begin{equation}
  S^{\mu\rho\tau}=-\frac{1}{\mu_0}(F^{\mu\rho}A^\tau-F^{\mu\tau}A^\rho)
  \label{eq:SAMT}
\end{equation}
that is easily found by explicitly evaluating the contribution due to $\delta_1
A^\lambda$ missing in (\ref{eq:Noether2}). One also easily finds that
\begin{equation}
  \partial_\mu S^{\mu\rho\tau}
  = -\frac{1}{\mu_0}\left( F^{\mu\rho} \partial_\mu A^\tau -F^{\mu\tau} \partial_\mu A^\rho \right)
  \neq 0~,
\end{equation}
where $\partial_\mu F^{\mu\sigma}=0$ was used since we consider here the
source-free case. This means that this term would contribute to the Noether
current for rotations. The corresponding current is determined by the
conservation law
\begin{equation}
  \partial_\mu \left(S^{\mu\rho\tau}
  -T_c^{\mu\nu} \op{X}_\nu^{~\rho\tau}\right) =0~,
  \label{eq:noether6}
\end{equation}
which follows from Eq.~(\ref{eq:noether6}) for rotations in the usual way by
Gauss's theorem. In other words, without restoring and adding the contributions
of $\delta_1 A^\lambda$, any results obtained for rotations would be incomplete
and meaningless.


\subsection{Belinfante symmetrization}\label{sec:Belinfante}

The Belinfante procedure~\cite{belinfante} seeks to bypass these problems by
providing a recipe for symmetrizing $T_c^{\mu\nu}$. The construction consists
of taking a linear combination of three spin-angular-momentum tensors
(\ref{eq:SAMT}),
\begin{equation}
  B^{\mu\lambda\nu} = -\frac{1}{2} \left(S^{\mu\lambda\nu} + S^{\nu\lambda\mu}-S^{\lambda\mu\nu}\right)~,
\end{equation}
and adding its four divergence to the canonical energy-momentum tensor
producing a new tensor,
\begin{equation}
  \EMT^{\mu\nu} = T_c^{\mu\nu} + \partial_\lambda B^{\mu\lambda\nu}~.
\end{equation}
Because the spin-angular-momentum tensors $S^{\mu\lambda\nu}$ are antisymmetric
in the last two indices, the Belinfante tensor $B$ is antisymmetric in the
first two indices, producing
\begin{equation}
  \partial_\mu \partial_\lambda B^{\mu\lambda\nu} = 0
\end{equation}
as a matter of course because the double contraction of a symmetric tensor
($\partial_\mu
\partial_\lambda$) with an antisymmetric tensor ($B^{\mu\lambda\nu}$) always
vanishes. Hence, both $\EMT^{\mu\nu}$ and $T_c^{\mu\nu}$ have the same
four-divergence,
\begin{equation}
  \partial_\mu \EMT^{\mu\nu} =  \partial_\mu T_c^{\mu\nu}~,
\end{equation}
which is essential for being able to extract the simplest possible conservation
law, namely four-momentum conservation following from translational invariance
equivalent to the basic assumption of homogeneity of space and time.

The Belinfante tensor evaluates here to
\begin{equation}
B^{\mu\lambda\nu} = \frac{1}{\mu_0} F^{\mu\lambda} A^\nu~,
\end{equation}
and thus
\begin{equation}
\partial_\lambda B^{\mu\lambda\nu} = \frac{1}{\mu_0}  F^{\mu\lambda} \partial_\lambda A^\nu~,
\end{equation}
where $\partial_\lambda F^{\mu\lambda} =0$ was used again. Adding this term to
$T_c^{\mu\nu}$ of Eq.~(\ref{eq:EMTunsym}), the new tensor then becomes
\begin{align}
  \EMT^{\mu\nu}
  &= -\frac{1}{\mu_0} F^\mu_{~\lambda} \partial^\nu A^\lambda-g^{\mu\nu}\mscrL
  +\frac{1}{\mu_0}F^{\mu\lambda} \partial_\lambda A^\nu
  \nonumber\\
  &=\frac{1}{\mu_0} F^\mu_{~\lambda} F^{\lambda\nu} -g^{\mu\nu}\mscrL~,
  \label{eq:EMTbelifante}
\end{align}
and it is indeed symmetric. To be viable, one still must show that the
contribution of the Belinfante tensor compensates the contribution of the
spin-angular-momentum tensor, and indeed it does since
\begin{equation}
  \partial_\mu \left(S^{\mu\rho\tau}
  +\partial_\lambda B^{\mu\lambda\nu}  \op{X}_\nu^{~\rho\tau}\right) =0
  \label{eq:noether7}
\end{equation}
and
\begin{equation}
  \partial_\mu \EMT^{\mu\nu}   \op{X}_\nu^{~\rho\tau}  =0
\end{equation}
are valid separately for rotations, the latter providing the --- now properly
constructed --- Noether-current tensor for rotations,  $M^{\mu\rho\tau}=
\EMT^{\mu\nu} \op{X}_\nu^{~\rho\tau}$.


\subsection{Accounting for gauge invariance}

All results presented so far are based on Taylor expansions in terms of
spacetime variations in $\delta x^\mu$. However, as was pointed out by
Bessel-Hagen~\cite{BesselHagen}, the Noether formalism~\cite{noether} permits
accounting for symmetries other than simple spacetime symmetries. Most notable
in this respect is the invariance of electrodynamics under gauge
transformations,
\begin{equation}
  A'^\nu(x) = A^\nu(x)-\partial^\nu \phi(x)~,
  \label{eqLgaugetrafo}
\end{equation}
where $\phi$ is a scalar function.

Given the fact that the underlying Lagrangian density (\ref{eq:Lagrangian}) is
manifestly gauge invariant since the field-strength tensor
(\ref{eq:fieldstrength}) trivially possesses this property, we now seek to make
this also a manifest property of the variational approach. To start, we
anticipate that we may break down the infinitesimal field transformation
(\ref{eq:barA}) into two contributions,
\begin{equation}
  \delta \bar{A}^\lambda = \delta_x A^\lambda +\delta_g A^\lambda~,
  \label{eq:deltaAall}
\end{equation}
where $\delta_x A^\lambda$ is a spacetime increment and the variation $\delta_g
A^\mu$ needs to account for  gauge transformations (\ref{eqLgaugetrafo}).  To
this end, let us employ the gauge transformation (\ref{eqLgaugetrafo}), go back
to Eq.~(\ref{eq:dAprime}),  and write it as
\begin{align}
  \delta \bar{A}^\lambda
  &= \left[ A^\lambda(x) -\partial^\lambda \phi(x) -A^\lambda(x') +\partial^\lambda \phi(x') \right]
  + \delta_1 A^\lambda
  \nonumber\\
  &=[A^\lambda(x)  -A^\lambda(x')] +\partial^\lambda \delta \phi
  + \delta_1 A^\lambda~,
\end{align}
where $\delta \phi=\phi(x')-\phi(x)$ is the infinitesimal gauge-function
increment. The first bracketed term here is precisely the spacetime increment
given by the right-hand side of Eq.~(\ref{eq:deltaAcanonical}),
\begin{equation}
\delta_x A^\lambda =-(\partial_\nu A^\lambda) \delta x^\nu~,
\end{equation}
that, by itself, produces the unsymmetric energy-momentum tensor in
Eq.~(\ref{eq:Noether2}). The remaining terms,
\begin{equation}
\delta_g A^\lambda = \partial^\lambda \delta \phi + \delta_1 A^\lambda~,
\end{equation}
then account for the implementation of gauge invariance.  Dimensional analysis
shows that the only infinitesimal scalar form linear in the field is
\begin{equation}
  \delta \phi = A_\nu \delta x^\nu~.
\end{equation}
Applying the product rule produces now
\begin{align}
\delta_g A^\lambda
&= \partial^\lambda A_\nu \delta x^\nu + \delta_1 A^\lambda
\nonumber\\
&=(\partial^\lambda A_\nu) \delta x^\nu +A_\nu \partial^\lambda \delta x^\nu  + \delta_1 A^\lambda~,
\end{align}
where
\begin{equation}
A_\nu \partial^\lambda \delta x^\nu =
A_\nu (\partial^\lambda \op{X}^{\nu\rho\tau})\omega_{\rho\tau} \equiv \delta_2 A^\lambda
\end{equation}
describes possible nonzero contributions from rotations if
$\omega_{\rho\tau}\neq0$, similar to $\delta_1 A^\lambda$ of
Eq.~(\ref{eq:delta1}). One easily finds that these contributions cancel
exactly,
\begin{equation}
\delta_1 A^\lambda+\delta_2 A^\lambda
=(A^\sigma \partial_\sigma \op{X}^{\lambda\rho\tau}+A_\nu \partial^\lambda \op{X}^{\nu\rho\tau})\omega_{\rho\tau}
=0~,
\label{eq:deltacancel}
\end{equation}
leaving the simple result
\begin{equation}
\delta_g A^\lambda = (\partial^\lambda A_\nu) \delta x^\nu
\end{equation}
for the gauge contribution. Hence, with (\ref{eq:deltaAall}), we obtain
\begin{equation}
  \delta \bar{A}^\lambda = -\left(\delta_\nu A^\lambda-\partial^\lambda A_\nu \right) \delta x^\nu
  =-F_\nu^{\,\lambda} \delta x^\nu~,
  \label{eq:dbarAF}
\end{equation}
which shows that the variation of $A^\lambda$ has now been made manifestly
gauge invariant --- and this, incidentally, also makes the separation of the
Euler--Lagrange equation (\ref{eq:ELeq}) from Eq.~(\ref{eq:Noether1})
manifestly gauge invariant.

In deriving this result, it is essential that the contributions arising from
derivatives of the spacetime increment $\delta x^\lambda$ for rotations cancel
completely, as shown in Eq.~(\ref{eq:deltacancel}), to avoid the problems for
the extraction of angular-momentum properties discussed in
Sec.~\ref{sec:roteffect} in connection with neglecting the rotation increment
$\delta_1 A^\lambda$ when defining the unsymmetric energy-momentum tensor from
Eq.~(\ref{eq:Noether2}).

Employing now this modification of $\delta\bar{A}^\lambda$ in the Noether
integral of Eq.~(\ref{eq:Noether0}) immediately results in the manifestly
gauge-invariant energy-momentum tensor,
\begin{align}
  \EMT^{\mu\nu}
  &=g^{\mu\sigma}\frac{\partial \mscrL}{\partial(\partial^\sigma A^\lambda) } F^{\nu\lambda}
-g^{\mu\nu}\mscrL
\nonumber\\
  &= \frac{1}{\mu_0}\left(g_{\beta\alpha} F^{\mu\beta}F^{\alpha\nu} + \frac{g^{\mu\nu}}{4} F^{\beta\alpha}F_{\beta\alpha}\right)~,
  \label{eq:EMT}
\end{align}
that is indeed manifestly symmetric, as expected. With its help, one can easily
extract all conservation laws of electrodynamics in the usual manner.


\section{Discussion}

The symmetric tensor (\ref{eq:EMT}) obtained here is identical to the one in
Eq.~(\ref{eq:EMTbelifante}) resulting from the Belinfante symmetrization.
However, its derivation is quite different since it directly employs all
symmetries inherent in the problem, including gauge symmetry, and, most
importantly, does not require an \emph{ad hoc} procedure like adding a
four-divergence with specific symmetry properties as in the Belinfante scheme
discussed in Sec.~\ref{sec:Belinfante}. Whereas the Belinfante symmetrization
recipe fixes something that is perceived to be flawed, the present approach
just employs all available variations applicable to the problem in a
straightforward application of Noether's Theorem.

Also, in deriving $\EMT^{\mu\nu}$ here, it was not necessary to make use of the
source-free equation $\partial_\mu F^{\mu\nu}=0$ anywhere. The tensor,
therefore, will also remain valid as a matter of course when adding an
interaction term to the Lagrangian, and thus can be utilized to derive more
complex conservation laws like energy and momentum conservation as summarized
in Poynting's theorems~\cite{jackson}. The usefulness of the source-free
energy-momentum tensor in such applications is usually tacitly assumed to be
the case anyway, but it is something that, strictly speaking, one needs to
verify.

For the case of source-free electrodynamics treated here, the derivation is
simple enough requiring only the modification (\ref{eq:deltaAall}) by adding
the gauge-transformation freedom to the usual spacetime variation. In view of
this, given the fact that source-free electrodynamics is conceivably one of the
simplest and most widely investigated gauge theories \emph{and} given the
additional fact that Noether's Theorem  is of such fundamental importance for
investigating symmetries and their related conservation laws, it seems quite
surprising that the full implementation of all symmetries for the problem, in
particular, gauge invariance, as a standard approach has not taken hold to this
day. This even more so since it seems to have been the intention of Noether to
have her formalism be understood that way, according to
Bessel-Hagen~\cite{BesselHagen}. In other words, to obtain an unsymmetric
energy-momentum tensor like the one in the integrand of Eq.~(\ref{eq:Noether2})
does not result from a `naive' application of Noether's Theorem, as it is
sometimes called~\cite{kaku}, but rather from an incomplete application.
Moreover, the derivation of the usual textbook result (\ref{eq:EMTunsym}) shows
that it relies on dropping essential rotational degrees of freedom which makes
the canonical tensor $T_c^{\mu\nu}$ only useful for the simplest application,
namely translational invariance. For anything more complicated, it is ill
defined and should not be used. This means, in particular, that suggestions to
settle the difference between canonical and symmetric energy-momentum tensors
experimentally are ill founded \cite{wang,afanasev}.

The presentation given here shows that in addition to the usual spacetime
transformations, gauge-invariance transformations are an essential additional
ingredient for providing all symmetries of source-free electrodynamics. In this
respect, it is interesting to note here that the two contributions arising from
the splitting (\ref{eq:deltaAall}) of $\delta\bar{A}^\lambda$, namely,
$\delta_x A^\lambda=-(\delta_\nu A^\lambda) \delta x^\nu$ and $\delta_g
A^\lambda=(\partial^\lambda A_\nu) \delta x^\nu$, both have a different origin.
One is an ordinary spacetime increment and the other the gauge-transformation
contribution, but both conspire to provide the gauge-invariant tensor
$F_\nu^{\,\lambda}$ making the final result (\ref{eq:dbarAF}) a mixture of
coordinate and field transformations.

The work of Bessel-Hagen clarifies that a complete implementation can be done
for all manner of gauge theories. Even though presented over a century ago, it
is only slowly gaining ground~\cite{eriksen,
munoz,burgess,montesinos,scheck,BLS,hobson}. It is hoped that the present
application to the simple case of source-free electrodynamics provides a
convincing example that the full implementation of symmetries in Noether's
Theorem for other applications may help avoid problems along the lines
encountered for electrodynamics. A case in point is the recent work of
Ref.~\cite{hobson} that applies the Bessel-Hagen procedure to a variational
treatment of gauge theories of gravity in a comprehensive manner.


\end{document}